\documentstyle[12pt]{article}
\title{
{\normalsize
\begin{flushright}
ITEP-31-95 \\
FTUV/95-18 \\
IFIC/95-18 \\
TPI-MINN-95/12-T \\
UMN-TH-1342-95 \\
\end{flushright}}
\vspace{0.3in}
How  strong can the coupling of leptonic photons be?\\}
\author{S.I.Blinnikov \thanks{Visitor at UCO/Lick Observatory,
    University of California, Santa Cruz, CA 95064},
A.D.Dolgov \thanks{Visitor at Institut de Fisica Corpuscular -
C.S.I.C. - Univ. de Valencia, Dept. de Fisica Teorica,
46100 Burjassot, Valencia, Spain},
L.B.Okun, and
M.B.Voloshin \thanks{Also: Theoretical Physics Institute, University of
Minnesota, Minneapolis, MN 55455, USA}\\
ITEP, Moscow, 117259, Russia\\}
\date{May 25, 1995}

\def\fun#1#2{\lower3.6pt\vbox{\baselineskip0pt\lineskip.9pt
\ialign{$\mathsurround=0pt#1\hfil##\hfil$\crcr#2\crcr\sim\crcr}}}

\begin{document}
\begin{titlepage}
\maketitle
\begin{abstract}
Consequences of possible existence of leptonic photon are considered for a
range of values of leptonic charge. In the case of a strong Coulomb-like
leptonic repulsion between electrons the existence of ordinary condensed
matter is impossible: antineutrinos cannot neutralize this destructive
repulsion.  The upper limit of leptonic charge is inferred from the
E\"{o}tv\"os type experiments. If however there exist light stable scalar
bosons with leptonic charge (e.g. singlet antisneutrinos) they may
neutralize the electron repulsion.  Possible experimental manifestations of
such leptonic bosons in gases and condensed matter are briefly discussed.
\end{abstract}
\end{titlepage}

\section{Introduction}

A possible existence of leptonic photons coupled to leptonic charge $e_l$
analogously to the coupling of the ordinary photons to the electric charge
$e$ has been discussed a long time ago \cite{1} following a paper of Lee and
Yang \cite{1a} on hypothetical baryonic photons. It was shown (see also
\cite{2}), that the coupling $\alpha_l = e^2_l/4\pi$ must be very small
($\alpha_l < 10^{-49}$ as compared to $\alpha = 1/137$) in order to comply
with the results of the experiments  \cite{3} - \cite{5} testing the
independence of gravitational acceleration on atomic number of the material
of which a pendulum is made of.  This  upper limit was derived in \cite{1}
assuming that the neutralization of leptonic charge of the Earth and the Sun
by the electronic antineutrinos is negligible.

A few years later $\alpha_l$  of the order of $10^{-8} - 10^{-9}$ was
postulated by arguing that the neutralization of the Earth's  leptonic
charge by $\bar{\nu}_e$'s is almost complete \cite{6},\cite{7}.
This claim was critically analyzed in ref. \cite{8}.

First, it was noted that $\alpha_l \leq 10^{-11}$ from the experiments on
$\nu_e e$-scattering. (This had been pointed out to the author of
ref. \cite{8} by  V.B.Geshkenbein and  B.L.Ioffe. At present this
experimental limit must be at least an order of magnitude better).

Second, it was argued that due to leptonic repulsion between atoms (because
of leptonic charge of the latter) the ordinary matter would be unstable
and a process of granulation would take place. Each granule would present a
kind of an ``atom", the nucleus of which would consist of a grain of
ordinary matter and the  ``atomic shell" -- of electronic
antineutrinos. The maximal, critical radius $r_c$ of a grain
would be determined by equilibrium between the leptonic repulsion,
proportional to $r^3$, and ordinary chemical short range attraction,
proportional to $r^2$.

Let us denote $N_c$ the number of ordinary atoms in a grain of a
critical size. Then
\begin{equation}
 \frac{(ZN_c)^2 \alpha_l}{(aN_c^{1/3})^2} \sim \frac{\alpha^2 \cdot
 N_c^{2/3}} {a^2}
 \label{1}
 \end{equation}
 where $a$ is the size of an ordinary atom, $Z$ is the
 number of electrons in it,  $r_c \sim aN_c^{1/3}$, and
 $\alpha^2/a^2$ stands for a crude estimate of the Van-der-Waals
 force; as an upper limit for a chemical force one can use
 $\alpha/a^2$. (In order to derive Eq. (\ref{1}) consider two halves
 of a grain of radius $r_c$. The l.h.s of Eq. (\ref{1}) is the force
 of Coulomb-like leptonic repulsion of the two halves, while the
 r.h.s is the chemical force of their surface attraction.)
 From Eq. (\ref{1}) it follows that a grain is stable if   $N
 < N_c$, where
 \begin{equation}
 N_c\sim \left(\frac{\alpha^2}{Z^2
 \alpha_l}\right)^{3/2}\;\;,
 \label{2}
 \end{equation}
 With $Z \sim 10$, $\alpha_l \sim 10^{-12}$, one gets $N_c\sim 10^9$
 and $r_c \sim 10^{-5}$ cm.  For $m_{\bar\nu_e} = 1$ eV the
 radius of the inner ``atomic" shell would then be $1/(ZN \alpha_l
 m_{\bar{\nu}_e}) \sim 10^{-3}$ cm while that of the outer shell
 $1/(\alpha_l m_{\bar{\nu}_e}) \sim 10^7$ cm.  Note that even the
 inner shell is much larger than the ``nucleus" and inside the
 nucleus there is practically no neutralization of the leptonic
 charge of atoms by antineutrinos.

 Third, ref. \cite{8} contained a simple argument due to Ya.B.Zeldovich,
 that an $\bar{\nu}_e$ cloud neutralizing electrons in the Earth
 cannot be  in equilibrium. Due to the Pauli principle, the momentum
 $p$ of $\bar{\nu}_e$  must be tens of keV
  (if their average density is equal to the electron density),
 $(p \sim Z/a$, where $a \sim 10^{-8}$ cm is the diameter of
 an atom and $Z$ - the number of electrons in it (for the Earth $Z
 \sim 20$)).  For a light $\bar{\nu}_e$ with $m_{\bar{\nu}_e} <
 100$ eV the kinetic energy is $E_{\rm kin} = p$. At the same time the
 average potential energy $U$ of attraction to an atom is $
 Z\alpha_l/a = Z(\alpha_l/\alpha) {\rm Ry} \sim 10^{-8}$ eV. Thus,
 $T >> |U|$ and the macroscopic condensed bodies are explosive.
In fact it is better to say
that in normal conditions they do not form because the
binding energy, $\sim 10^{-8}$ eV, is much smaller than the
characteristic energy of neutrinos.  The
 system is more like an electromagnetic plasma at temperatures higher
than the hydrogen recombination temperature. We
consider the mechanism of instability in some detail in Sec.2.

The final remark of ref. \cite{8} referred to a hypothetical situation when
the leptonic charge of atoms is neutralized by a special kind of
leptonically charged bosons. In this case the leptonic charge of the earth
could be compensated. ``It is interesting to find out at what values of
$\alpha_l$ and of the lepton boson mass, the existence of the condensate of
such bosons does not contradict the established physical, chemical and
mechanical properties of the matter" \cite{8}.  The bosonic case which was
recently considered in ref. \cite{9} is discussed in  Sec. 3.

\section{Condensed matter cannot be
neutralized by antineutrinos}

Let us consider neutralization by $\bar\nu_e$
of the leptonic charge of
electrons in a macroscopic body with the density of electrons
$n_e(\vec{x})$. We start with assuming that the system is in
a state of equilibrium and will prove that this assumption is wrong.

The profile of the antineutrino density
$n_{\bar{\nu}}(\vec{x})$ can be found by means of the Thomas-Fermi
equation, which in this case is derived as follows. Let
$\phi(\vec{x})$ be the potential energy of one antineutrino in the
potential of the field associated with the leptonic charge. The
behavior of $\phi(\vec{x})$ is determined by the Poisson equation
\begin{equation}
\Delta\phi = 4\pi\alpha_l(n_e -n_{\bar{\nu}})
\label{poisson}
\end{equation}
For a degenerate
antineutrino gas the density $n_{\bar{\nu}}$ is given by the
Fermi momentum $p_F$ of the $\bar{\nu}_e$ as \footnote{We assume that
$\bar{\nu}_e$ is two-component. The assumption of the degeneracy will
be discussed below.}
\begin{equation}
n_{\bar{\nu}}=\frac{1}{6\pi^2}
p_F^3 \;\; .  \label{fermi}
\end{equation}
The condition of equilibrium for the
antineutrino gas is that the sum of the kinetic and the
potential energy should be a constant, independent of $\vec{x}$. For
a relativistic antineutrino, $p_F\gg m_{\bar{\nu}}$, the latter
has the form
\begin{equation} p_F +\phi =c \;\; .
\label{energy}
\end{equation}

The constant $c$ is fixed by the boundary condition at infinity:
$c=0$. Thus the Poisson equation (\ref{poisson}) can be rewritten in
a closed form in terms of the position-dependent Fermi momentum
$p_F$:
\begin{equation}
\Delta p_F = -4\pi\alpha_l\left(n_e -\frac{1}{6\pi^2} p_F^3\right)
\label{thomas}
\end{equation}

If the profile $n_e(\vec{x})$ of the electron density is parametrized in
terms of a characteristic value $n_0$ as $n_e(x)=n_0 f(\vec{x})$ with
dimensionless function $f(\vec{x})$, the equation (\ref{thomas}) can be
rendered in the form
\begin{equation}
\Delta u(\vec{x}) =
\mu^2
(f(\vec{x})-u^3(\vec x)) \;\; ,
\label{thomas-u}
\end{equation}
where $u(\vec{x})=p_F(\vec{x})/(6\pi^2 n_0)^{1/3}$ is the
dimensionless ratio of the momentum $p_F$ for neutrinos to an
effective Fermi momentum, characterizing the electron density $n_0$,
and
\begin{equation}
\mu = \biggl(\frac{4\pi\alpha_l}{(6\pi^2)^{1/3}}\biggr)^{1/2} n_0^{1/3}
\label{parameter}
\end{equation}
is the corresponding ``mass parameter".

In order to estimate the magnitude of the quantities involved and
also to justify the assumptions about the degenerate gas of
relativistic antineutrinos, let us consider a solid body of
characteristic size $L$, with a uniform density $n_e =n_0\approx
10^{25} {\rm cm}^{-3}$, which corresponds to
an effective Fermi momentum $p_0
=(6\pi^2 n_0)^{1/3}\sim 1.7\cdot 10^4$ eV, which is much larger than
both the possible neutrino mass and a possible temperature in
terrestrial conditions. The mass parameter $\mu$ then has the
characteristic value $\mu\simeq 8\cdot
10^{-3}\sqrt{\alpha_l/10^{-12}} {\rm eV} \simeq \sqrt{\alpha_l/
 10^{-12}} (3\cdot 10^{-3} {\rm cm})^{-1}$. In this model example the
function $f(\vec{x})$ is $f(\vec{x}) =1$ inside the body and
$f(\vec{x}) =0$ outside the body. If the size of the body $L$ is
large, $L\gg 1/\mu$, the solution of the equation
exponentially approaches $u=1$ from the surface inwards, the scale in
the exponent being set by the skin depth $1/\mu$. Therefore, the
leptonic charge of the bulk of the body is completely neutralized
($u=1$), except for the skin layer. Outside the body the solution for
$u(\vec{x})$ dies down as a power of distance, if vacuum is assumed at
infinity.
(More precisely, one should
expect an  exponential Debye screening by the sea of
cosmic antineutrinos, even if they are massless.
The screening length is
$D \sim (E_{\bar{\nu}} / \alpha_l n_{\bar{\nu}})^{1/2}$ where
$E_{\bar{\nu}}$ is
the characteristic antineutrino energy and $n_{\bar{\nu}}$ is their
number density.  For the average cosmological values $n \simeq 100\;
{\rm cm}^{-3}$ and $E\simeq 2$ K the screening length is $D\simeq
10^5$ cm.  By the way, this means that the outer shell of an ``atom"
considered in Introduction should be much smaller due to the Debye
screening.
For massive nonrelativistic neutrinos the galactic number density may
be larger and kinetic energy $E$ smaller, so correspondingly the
screening length $D$ would be considerably smaller.
The result depends
upon velocity distribution and the neutrino mass
(for more details see Sec. 3).
For the neutrino gas with the parameters
considered in this Section the results qualitatively
do not change due to large Debye
length $D \sim 1 / e_l E_{\bar{\nu}}$.)

Thus,
one is tempted to conclude that the bulk leptonic charge of
macroscopic objects could be neutralized, unless the dimension $L$ is
small in comparison with the skin depth, $L\ll 1/\mu$. (One can
readily see that for objects whose all three dimensions are of the
same order, the latter condition is equivalent to an upper bound on
the total number of electrons $N_e$ in the object: $N_e \ll
\alpha_l^{-3/2}$. For such objects the
antineutrino density inside them
is small: $n_{\bar{\nu}}\ll n_e$, so that local neutralization of the
leptonic charge does not take place as was already discussed in the
Introduction).

Let us now consider large objects with  $L \gg 1/\mu$.
In the previous discussion it was assumed that the electron density
$n_e(\vec{x})$ was fixed by the electron binding in atoms and the
overall stability of the body was ensured by the elastic forces,
i.e.  the back reaction of the pressure of the antineutrino gas on
the stability of matter was ignored. We will show now that for
ordinary substances this assumption is in fact violated by many
orders of magnitude independently of $\alpha_l$, provided that the
condition $L\gg 1/\mu$ is satisfied. Indeed, in the
bulk of the body $n_{\bar{\nu}}=n_e$, while exactly on the surface
$(n_e -n_{\bar{\nu}})/n_e = O(1)$.  For  the pressure of the
antineutrino gas inside the body we have
\begin{equation}
P \simeq
 n^{4/3}_{\bar{\nu}} \simeq  n^{4/3}_e \; ,
\label{pressureprime}
\end{equation}
Therefore, the skin layer experiences the
pressure difference of the order of $n_0^{4/3}\approx 7\cdot 10^{15}$
Pa. The highest pressure sustained by known materials is about
$1.5\cdot 10^9$ Pa, while for most solids it is typically $10^7 -
10^8$ Pa, not even to mention liquids. Thus it is impossible to
stabilize the neutrino pressure by elasticity of materials.

There is another, more simple way to derive the same results. Let us
consider the ``capacitor" formed by atoms and $\bar{\nu}_e$ in the
skin layer. The pressure $P$ of the degenerate antineutrino gas
is balanced by the leptonic attraction between oppositely charged
layers with the surface leptonic charge density $\sigma =e_l
n_{\bar{\nu}}b$, where $b$ is the thickness of the skin. Then the
attraction force per unit area is \begin{equation}
\frac{F}{S}=4\pi\sigma^2 =4\pi\alpha_l n_{\bar{\nu}}^2 b^2
\label{capacitor}
\end{equation}
On the other hand, $F/S=P$ with $P = p_F^4/24\pi^2$ and
$n_{\bar{\nu}}= p_F^3/6\pi^2$.  From this we get
\begin{equation}
p_F^4/24\pi^2 =4\pi\alpha_l\left(\frac{p_F^3}{6\pi^2}\right)^2 b^2
\label{capacitor energy}
\end{equation}
\begin{equation}
 b={1 \over p_F} \sqrt{{{3\pi} \over 8\alpha_l }}   =
{1 \over 4} \left ( {6 \over \pi} \right )^{1/6}\, n_e^{-1/3}\,
\alpha_l^{-1/2}~,
\label{be}
\end{equation}
which is very close to the previous estimate of skin thickness
$1/\mu$ with $\mu$ from  Eq.(\ref{parameter}) (differs by factor 2).

 Note that if we rewrite $n_e$ in terms of the Bohr radius
$a_0 \simeq 0.5
 \cdot 10^{-8}$ cm,
 \begin{equation}
 n_e \simeq \frac{3Z}{4\pi a^3_0}\;\;,
 \label{Bohr}
 \end{equation}
 then
 \begin{equation}
 b \sim \frac{a_0}{\sqrt{\alpha_l}} Z^{-1/3}
 \label{again}
 \end{equation}
 For $Z=10$ and $\alpha_l = 10^{-12}$ we find $ b \simeq 10^{-3}$ cm.

Therefore  the source of nonequilibrium is the thin skin layer. The
electrons in the skin leptonically repel each other, so that parts of the
skin are explosively peeled off.  This process would continue till the whole
body would blow up producing small pieces of dust with the size evaluated
in the Introduction. For more compact bodies gravitational attraction might
dominate like that in white  dwarfs, where gravity compensates Coulomb
repulsion of ions.

By decreasing $\alpha_l$ we  get an increased thickness of the skin. But
the peeling pressure
 \begin{equation}
 P \simeq \left(\frac{a_0}{\sqrt{\alpha_l}}\right)^2
\cdot \frac{\alpha_l}{Z^{2/3}}
 n^2_e \sim \frac{Z^{4/3}}{a^4_0}
 \label{pressure}
 \end{equation}
remains constant as long as $L\mu\gg 1$.

The atomic forces are thus insufficient to compensate the repulsion of
electrons.  Therefore one has to seek for some outer force to keep the
system bound.  The only force conceivable in terrestrial conditions is the
gravity, it gives the binding energy around $\alpha_N N_\oplus/R_\oplus$ per
electron where $\alpha_N = G_N m_b^2 \sim 10^{-38}$, and $N_\oplus\sim
10^{51}$ is the number  of baryons in the Earth (which is approximately
twice the number of electrons), $R_\oplus \simeq 10^9$ cm  being the Earth
radius.  (We substitute the baryon mass $m_b$ in this estimate and not
$m_e$, since actually the gravity binds baryons whereas electrons are
coupled to them by ordinary electric field - cf. below for the case of
bosons).  We find that the gravitational binding energy is four orders of
magnitude lower than  required.  Thus antineutrinos with density
$n_{\bar{\nu}}=n_e$  would not stay inside the Earth even if by some miracle
they had been pumped in.

As the electrons in the Earth and in the Sun cannot be neutralized by
antineutrinos the upper limit on $\alpha_l$ is valid, derived in ref.
\cite{1} on the basis of the E\"{o}tv\"{o}s-type experiments.

\section{Sneutrinos in the atmosphere and in condensed matter}

Let us consider now the case of bosonic leptons, e.g.
$\tilde{\bar{\nu}}$: antisneutrinos
\footnote {$\tilde{\bar{\nu}}$ --
superpartner of the left-handed ``sterile" antineutrino, which does
not participate in the electroweak interactions; otherwise it would
manifest itself in the decays $Z \to \tilde{\bar{\nu}}\tilde{\nu}$
and hence in the invisible width of the $Z$ boson.},
assuming that their mass is very
small:  of the order of 1 eV.
This
possibility was raised in the paper \cite{9}.

The authors of \cite{9} suggest to estimate the effect of the
compensation by using the same Poisson equation  (\ref{poisson}),
\begin{equation}
   \Delta\phi(\vec x) = 4\pi \alpha_l
        \bigl(n_e(\vec x)-n_l(\vec x)\bigr)  \; ,
  \label{eq:poiss}
\end{equation}
where $n_l$ denotes the number density
of the sneutrinos $\tilde{\bar\nu}$ the
distribution of which
was assumed to obey the Boltzmann law
\begin{equation}
        n_l(\vec x) = n_0\exp(-\phi(\vec x)/T) \; .
  \label{eq:boltz}
\end{equation}
The equations (\ref{eq:poiss}) and (\ref{eq:boltz})  are
written in analogy with electrodynamics of gases obeying Boltzmann
statistics \cite{10,LL}. If we pursue this analogy for the gas of
scalar particles with temperature $T$ then we can estimate a
characteristic  length of screening
\begin{equation}
        D \sim  (T/\alpha_l n_l )^{1/2} \; ,
  \label{eq:deb}
\end{equation}
This is  the  well-known Debye radius,
defining the range of the screened potential of
an electron (see the discussion after  Eq. (\ref{parameter}) above,
cf.  e.g.  \cite{10,LL} for the case of classical plasma)
\begin{equation}
\phi(r) \sim \exp(-r/D)/r \; .
\label{eq:yukpot}
\end{equation}
The typical distance of interaction of leptons
colliding with an atom with $Z$ electrons can be estimated as
\begin{equation} r_0 \sim \alpha_l Z/T \; .  \label{eq:dist}
\end{equation} This gives the cross-section
\begin{equation}
\sigma \sim r_0^2 \Lambda \sim \alpha_l^2 Z^2\Lambda /T^2 \; ,
       \label{eq:sigm}
       \end{equation}
with $\Lambda$ being the
Coulomb logarithm, $\Lambda \sim \ln(D/r_0)$.  Now the time of energy
relaxation (or rather of temperature equilibration for
$\tilde{\bar\nu}$ and ordinary gas) is easily found:
\begin{equation}
   t_T = {m_a \over m_l n_l \sigma v_l }
   \simeq  {m_a T^{3/2} \over \alpha_l^2 Z^2 m_l^{1/2} n_l \Lambda}
   \; ,
   \label{eq:trel}
   \end{equation}
   where $m_a$ is the mass of an  atom.
   The relaxation time
$t_T$ is the order of years for $\alpha_l \simeq 10^{-12}$, $m_l
\simeq 1$ eV, \quad $n_l \sim 10^{20}$ cm$^{-3}$ and $T\simeq 300$ K.

If we assume that the lepton charge is compensated on scales larger
than $D$ in the thermally relaxed Boltzmann gases of $\tilde{\bar\nu}$'s
and of the air in the
terrestrial atmosphere then we can show that inevitably a large field $E_l$
appears. This field is necessary to support an approximate equality of
concentrations of $\tilde{\bar\nu}$'s and of electrons.
Everything is quite analogous to
the charge neutrality of plasma in stellar atmospheres which is always
slightly violated in the presence of non-electric forces \cite{pikl}.
Let us use the equations of equilibrium
\begin{equation}
   \nabla P_a = - n_a m_a g + n_a Z e_l E_l  \; ,
  \label{eq:grpa}
\end{equation}
and
\begin{equation}
\nabla P_l = - n_lm_l g - n_l e_l E_l \;.
  \label{eq:grpl}
\end{equation}
where $P_a$ and $P_l$ denote  pressures of atoms and
leptons respectively
and $g$ is the acceleration
produced by gravity. We will show now that one cannot satisfy
both equations (\ref{eq:grpa}) and  (\ref{eq:grpl}) if $E_l=0$,
i.e.  assuming ideal compensation of charges $Zn_a = n_l$, since the
gravitational force on atoms is much larger. Yet remembering that the
equality $Zn_a \simeq n_l$ is only approximate (though it may be very
accurate in fact), one sees that this slight disbalance of charge
densities produces non-negligible field $E_l$. Assuming equal
temperatures of atoms and of $\tilde{\bar\nu}$'s we then find
\begin{equation} m_a g - Z e_l E_l \simeq m_l g + e_l E_l  ,
   \label{eq:forc} \end{equation} since the pressure of Boltzmann
atoms is $P_a=n_a T$, and the pressure of Boltzmann
$\tilde{\bar\nu}$'s   is $P_l =n_l  T$.  Hence, \begin{equation} e_l
   E_l  \simeq  m_a g / (Z+1).  \label{eq:field} \end{equation}

Thus the new force acting on atoms
would be comparable to the force of
gravity (irrespective of the value of $\alpha_l$ if it is large enough to
provide for the
dynamic equilibrium).
This strong effect could be detected on the bodies with
differing ratio $AZ/(Z+1)$, i.e. already in the Newton experiments
to say nothing on E\"otv\"os-Dicke
experiments. This means that something is wrong in our basic
assumptions: either the thermal equilibrium does not hold (due to the low
value of $\alpha_l$), or the
Boltzmann approximation to the distribution function is not valid.
One should also consider the
neutralization of leptonic charge produced by
$\tilde{\bar\nu}$'s
concentrating in the condensed bodies, since the estimate
(\ref{eq:field}) is directly applicable only to the atoms in the air.

First of all, we have to take into account
that the sneutrinos are bosons. Let us
estimate the conditions under which the gas consisting
of these particles becomes degenerate,
that is when the Bose condensate forms.
We can find the temperature of condensation in the following way.

The number density of bosons
\begin{equation}
n = \int f d^3 p / ( 2\pi )^3
 \label{nint}
\end{equation}
where $f$ is the occupation number
\begin{equation}
f =  1/[\exp ((E-\mu)/T ) - 1 ]
 \label{fnorm}
\end{equation}
for the thermal equilibrium with temperature $T$ and chemical potential
$\mu$. In the presence of the condensate the expression for $f$ is modified:
\begin{equation}
f = (2\pi)^3 C \delta^3 (p) + 1/[\exp ((E-m)/T ) - 1 ] \; .
 \label{fcond}
\end{equation}
The first term corresponds to the particles in the condensate and the
second is the normal Bose term with chemical potential equal to the mass.
The coefficient $C$ is found from the condition that the particle number
density is
\begin{equation}
n = n_l \; .
 \label{nact}
\end{equation}
If the first term in the expression for $f$ is the dominant one,
then we can conclude that $C=n_l$. In the case of the Earth's atmosphere
$n = n_l = n_e$, where $n_e$ is the number density of electrons.
To find the total number of particles
we have to calculate the integral
\begin{equation}
 n = \int {d^3 p \over (2\pi)^3} \left[ \exp \left( {E-\mu\over T}
\right) - 1 \right]^{-1}  \; .
 \label{n}
\end{equation}
The number density of the particles outside the condensate, $ n_{\rm out}$,
is given by Eq.(\ref{n}) with $\mu=m$.

Introducing
\begin{equation}
  u = {E - m \over T}\;, \qquad \beta={m \over T}
 \label{newvar}
\end{equation}
we rewrite (\ref{n}) as
\begin{equation}
n={m^3 \over 2\pi^2\beta^{3/2}} \int_0^\infty\!
{(1+u/\beta)(2+u/\beta)^{1/2}
 u^{1/2} \over \exp [u-\beta(\mu/m-1)] - 1}\, du\;.
\label{iripa}
\end{equation}
Now, for $T\ll m$, i.e. $\beta\gg 1$, we find the number density
of bosons outside the condensate putting $\mu=m$ in (\ref{iripa}):
\begin{equation}
n_{\rm out}={m^3 \over 2^{1/2}\pi^2\beta^{3/2}} \int_0^\infty\!
 {u^{1/2} du \over \exp u - 1}\; =
 \left({mT\over 2\pi}\right)^{3/2}\zeta(3/2)  \; ,
\label{nonrel}
\end{equation}
(see e.g. [16]).

For high temperature, $T\gg m$, i.e. $\beta\ll 1$,  we find from
(\ref{iripa})
\begin{equation}
n={m^3 \over 2\pi^2\beta^3} \int_0^\infty\!
{ u^2 du \over \exp [u-\beta(\mu/m-1)] - 1}\;.
\label{prel}
\end{equation}
and for $\mu=m$
\begin{equation}
  n = T_c^3\zeta(3)/\pi^2 \; .
\label{corncond}
\end{equation}
Thus, keeping in mind that in the Earth's atmosphere at
the sea level there are $2.7\times 10^{19}$ molecules/cm$^3$, i.e.
the electron concentration is $n_e \simeq 4\times 10^{20}$, we get
for the temperature of condensation
\begin{equation}
  T_c=(\pi^2 n_l/\zeta(3))^{1/3} \approx 30 \,{\rm eV} \; .
\label{cortcond}
\end{equation}
This can be compared to $T = 300 \,{\rm K} = 0.03$ eV.

In deriving (\ref{cortcond}) we have neglected processes of pair production
and annihilation.
In general, at high temperature one has to take into account the
presence of antiparticles. If the coupling of pairs to the leptonic photons
is strong enough and the dimensions of the system are large enough
(so that the antiparticles are trapped) then
the concentration of antiparticles $\bar n$ is found from (\ref{prel})
by substituting $-\mu$ instead of $\mu$, since the pairs are in
equilibrium with blackbody radiation, having zero chemical potential.
In this case the condensate is formed when the excess of particles over
antiparticles
is so big that the maximum possible value of $\mu = m$ is reached.
The excess of particles
over antiparticles in the limit of small $\beta$ and small $\mu/T$ is

\begin{equation}
\delta n= n-\bar n =
 {m^3 \over 2\pi^2\beta^3} \int_0^\infty\!
  u^2 \left({-2\mu \over T} \right) {d\over du}{1\over
  \exp u - 1} du
  = { \mu T^2 \over 3} \;.
 \label{excn}
\end{equation}

The lowest value of the temperature before the formation of condensate
can be determined from the equation
$\delta n = m\, T^2_c/3$. At low temperatures, $T\ll m$, the
number density of particles $n$ which survived the annihilation is equal
to $\delta n$, defined by Eq. (\ref{excn}).
Hence, for the Earth's atmosphere
\begin{equation}
T_c = \sqrt {3 n_l /m_l } \approx 3 \,{\rm keV}
\label{eq:tcond}
\end{equation}
for $n_l \simeq n_e$ and $m_l\simeq 1$ eV.

However, we cannot apply the expression (\ref{eq:tcond}), which is good
for cosmology, to terrestrial
conditions, or to conditions in the protoplanetary nebula.
For small values of $\alpha_l$
and for temperatures lower than few keV we cannot expect trapping of
antiparticles and the equilibrium of pairs with radiation. Here the
Eq.(\ref{cortcond}) is more relevant.

In any case for the small mass of
$\tilde{\bar\nu}$'s almost all of them enter the condensate. The
fraction of the rest is
$ ( mT / 2\pi )^{3/2} \zeta (3/2) / n_l \approx 10^{-9}$.
Their pressure depends only on $T$, and one cannot prove
the appearance of nonvanishing field $E_l$ using the same
arguments which have led to the result (\ref{eq:field}).

If the bosons were neutral, the condensate  would be collected just in
the center of the gravitational potential well \cite{tkach}, i.e. in the
Earth's center.
In our case one cannot claim this: it is clear that the mutual repulsion
of $\tilde{\bar\nu}$'s
prohibits their high spatial concentration.

What is the structure of the condensate? In a very rarefied gas the
$\tilde{\bar\nu}$'s should form `atoms' with electrons. The radii of these
atoms are by a factor of $(\alpha/\alpha_l)(m_e/m_l)$ larger than the Bohr
radius (see above), i.e. they are in the range from meters through thousand
kilometers for $\alpha/\alpha_l \geq 10^{10}$ and for  $m_l$ in the range
from $m_e$ down to 1 eV. In any case this is much larger than the mean
distance between electrons in the terrestrial conditions.  Using the
arguments of Feynman \cite{feynstat} one can convince oneself that the
condensate, i.e. the ground collective state of the scalar particles, must
be a quasi-uniform distribution in  a self-consistent field $ E_l$, created
by the charge density $n_e-n_l$ (where $n_l$ is the total density of
$\tilde{\bar\nu}$'s accounting also for the particles outside the
condensate.

Let us compare now the problem of the skin layer for bosons with the fermion
case, considered in Sec.2.  In the boson  case, the thickness of the skin is
of the order of the Compton wave length of the boson:
$b=1/m_{\tilde{\bar{\nu}}}$, so that the force of the leptostatic attraction
between electrons and antisneutrinos per unit area (which is to be balanced
by the pressure due to elastic forces) at the
surface is $\alpha_l n_e^2 m^{-2}_{\tilde{\bar{\nu}}}$.  It is easy to see
that in this case the force is weaker than in the case of fermionic
antineutrinos and decreases with decreasing $\alpha_l$. The body itself is
filled with bosonic condensate of $\tilde{\bar{\nu}}$'s (at rest). Thus an
equilibrium can be reached and the allowed values of $\alpha_l$ depend on
$m_{\tilde{\bar{\nu}}}$.

Let us now estimate the field $E_l$, accounting for the
condensate, using equations (\ref{eq:grpa}) and (\ref{eq:grpl}).
In an isothermal atmosphere one finds a rather weak force
\begin{equation}
   e_l E_l = - m_l g \;,
  \label{eq:weakf}
\end{equation}
which is $m_a/m_l$ times smaller than the gravity for atoms. Given the
results of the E\"otv\"os type experiments one may derive a bound on
$m_l$ from this.

Consider another, more realistic case: let the atmosphere be
non-isother\-mal and the gravity in (\ref{eq:grpl}) can be neglected
(if the mass $m_l$ is sufficiently small). Then, since
$ \nabla P_a \simeq - n_a m_a g $ (the force produced by the field $E_l$
on atoms is negligible) and
$P_l \sim (T/T_c)^{3/2}(m_l/T_c)^{1/2} P_a$,
we find a crude estimate
\begin{equation}
   Ze_l  E_l \sim  (T/T_c)^{3/2}(m_l/T_c)^{1/2} m_a g \; .
  \label{eq:weakondf}
\end{equation}
For the Earth atmosphere we find the force of the same order (but of the
opposite sign) as for the isothermal case. In both cases one may not
assume the equality of the temperatures of atoms and scalars. The
estimate (\ref{eq:weakondf}) varies according to the temperature $T$
of the gas of
$\tilde{\bar\nu}$'s.

We find that the presence of the condensate strengthens the chances for
perfect compensation of charges. One should remember, however, that
in a real (non-ideal) Bose-gas, like superfluid helium, an
appreciable fraction of bosons remains outside the condensate even at
zero temperature.
Thus the actual estimate of the force must be somewhere
in between the Boltzmann and the ideal Bose estimates considered here.
The weakly excited condensate behaves like a superfluid classical
liquid and the flow of this liquid must be potential
\cite{feynstat,11}. So, even if one imagines that all macroscopic
bodies on the Earth surface are filled with a condensate which
screens the electron field $E_l$ inside them (one can imagine an
adiabatic process of formation of those bodies embedded in the
uniform cloud of ordinary matter filled with uniform condensate),
then one should encounter manifestations of the leptonic charge when
the motion of the bodies is not potential and non-adiabatic (like in
collisions, etc.).

Perhaps, it is best of all to consider the constraints on $\alpha_l$ on
small bodies, like dust grains, in the absence of the condensate in
the interstellar (or circumstellar) space, where the gas density is
almost vacuum (20 orders lower than in our atmosphere). Let us assume
that the constants are such as to give for the simplest electron-scalar
"atom" a radius $r_a$. If we have a dust grain with $ZN$
not screened electrons ($N$ is the number of atoms in the grain,
and $Z$ is number of electrons in an atom) then the
$\tilde{\bar\nu}$'s tend to occupy the ``atomic" orbits with radii
$r_a/ZN$.  If $ZN$ is very high, then the sneutrinos tend to screen
the electrons inside the grain,
but if $ZN$ is sufficiently small then all sneutrinos
remain outside the grain.  E.g. if $r_a=10^8$ cm and the
density of electrons in the grain is $10^{24}$ cm$^{-3}$, then
at $ZN \sim 10^{12}$ the radius of
a grain $r_g\sim 10^{-4}$ cm is of the order of the radius of ``atomic"
shell.  Note that this is an order of magnitude larger than the
critical radius $r_c$ at which the grain is disrupted by mutual
leptonic repulsion of unscreened electrons (see Introduction). Thus
there should be a gap in the spectrum of grain sizes at $r_g\sim
10^{-5} - 10^{-4}$ cm.

It is known \cite{18} that much smaller grains do exist in the interstellar
space.  There is no convincing evidence that there is a gap in the region of
$r_g\sim 10^{-5} - 10^{-4}$ cm, but some models of the interstellar
extinction do not exclude that the grain size distribution falls sharply
above $0.25\times10^{-4}$ cm (see references in \cite{18}).  This can have
quite natural explanation without invoking new physics because of
grain-grain collisions leading to grain fragmentation. Of course, larger
grains must exist and grow in protoplanetary clouds (otherwise, the
formation of planets would be impossible). Unfortunately, the theory of
evolution of interstellar dust is far from being complete \cite{19}.  It
would be interesting to explore this question with the new physics discussed
here in a broad range of $\alpha_l$, $m_l$ and to compare this with
observations of the interstellar dust.

\section{Concluding remarks}

Thus we have shown that the old upper limit \cite{1} on $\alpha_l$ is valid
if there is no leptonically charged scalar bosons. If such bosons exist, they
may screen the leptonic charge of electrons and the situation is less clear.
We made our estimates by assuming as reference values $\alpha_l = 10^{-12}$
and $m_l = 1$ eV. It is important to stress that the mass of sneutrinos,
$m_l$, may be much larger. It is very interesting to consider such phenomena
as surface tension, surface waves and capillarity (a remark by
S.G.Tikhodeev), as well as electrolysis, melting of metals, boiling of water
and other phase transitions in the presence of the condensate of leptonic
bosons. In this way it would be possible to establish the allowed region on
the plane $\alpha_l$, $m_l$.

\vspace{5mm}

{\bf Acknowledgements}

One of the authors (L.B.O.) is grateful to S.G.Tikhodeev for very useful
discussions and to the authors of ref. \cite{9} for giving him their
preprint during a Physics School at T\"{u}bitak, near Istanbul. S.I.B. is
grateful to D.Osterbrock and to D.Lin for helpful discussions on the problem
of interstellar dust.  The work of S.I.B. is supported in Russia in part by
the Russian Foundation for Fundamental Research grant No. 93-02-03637 and
in  the US by the National Science Foundation (AST-91-15367) and by NASA
(NAGW-2525). The work of M.B.V is supported in part by the DOE grant
DE-AC02-83ER40105.  The work of A.D.D. is supported by DGICYT under Grant
numbers PB92-0084 and SAB94-0089 (A.D.). The work of L.B.O. is supported by
RFFR grant No. 93-02-14431.


\newpage
\begin{thebibliography}{99}
\bibitem{1} L.B.Okun. Yadernaya Fizika {\bf 10} (1969) 358 (in
Russian). Sov.Journal of Nucl.Phys. {\bf 10} (1969) 206 (in
English).
\bibitem{1a} T.D.Lee and C.N.Yang, Phys. Rev. {\bf 98} (1955)
1501.
\bibitem{2} L.B.Okun. Leptons and Quarks. North Holland. 1982, p. 166.
\bibitem{3} R.V.E\"{o}tv\"{o}s, D.Pekar and E.Fekete, Ann. der Phys.
{\bf68} (1922) 11.
\bibitem{4} R.H. Dicke and P.G.Roll, G.Krotkov, Ann. of Phys.
{\bf26} (1964) 442.
\bibitem{5} V.B.Braginsky and V.I.Panov, Sov. Phys. JETP {\bf 34}
(1972) 463.
\bibitem{6} G.A. Zisman. Uchenye Zapiski LGPI  No. 386 (1971) 80 (in
Russian).
\bibitem{7} V.M.Goldman, G.A.Zisman, R.Ya.Shaulov. Tematicheskiy Sbornik
LGPI, 1972 (in Russian).
\bibitem{8} L.B.Okun. A remark on leptonic photons. June 1972 (unpublished,
in Russian, presented at the ITEP theoretical seminar).
\bibitem{9}
A.K.\c Cift\c ci, S.Sultansoi, \c S.T\"urk\"oz.
Lepton charge as possible source of new
force. Ankara University preprint AU/94-03(HEP).
\bibitem{10}R.P.Feynman,
R.B.Leighton, M.Sands, The Feynman Lectures on Physics. Addison-Wesley,
1963, v.1. Chapter 40, v.2 Chapter 7.
\bibitem {LL} L.D. Landau and E.M. Lifshitz. Statistical Physics,
Part 1, Nauka, Moscow, 1976.
\bibitem {pikl} S.B.Pikelner. Foundations of cosmic electrodynamics,
  Nauka,  Moscow, 1968. Chapter 1 (in Russian).
\bibitem {tkach} I.I.Tkachev. Phys.Lett. B261(1991)289.
\bibitem {feynstat} R.P.Feynman. Statistical Mechanics, W.A.Benjamin,
  Reading, 1972. Chapter 11.
\bibitem{11} L.D.Landau and E.M.Lifshits.
 Theoretical Physics v.IX.  E.M.Lifshits and L.P.Pitaevsky. Statistical
 Physics, part 2.  Theory of condensed state. Moscow Nauka, 1978. Chapter
 III (in Russian).
\bibitem{8a} L.D.Landau, E.M.Lifshitz. Quantum Mechanics.
Non-Relativistic Theory. Nauka, Moscow, 1989, Par. 70 (in Russian).
\bibitem{18} J.S.Mathis. Annual Review on Astronomy and Astrophysics,
{\bf 28} (1990) 37.
\bibitem{19} B.T.Draine {\em in} Astronomical Society of the Pacific
Conference Series. Vol. 12. The evolution of the interstellar medium.
Ed. L.Blitz, San Francisco, 1990, p.193.
\end{thebibliography}
\end{document}